\documentstyle[epsf,seceq,preprint]{ptptex}

\newcommand{\VSPACE}{\vspace{3ex}}
\newcommand{\wt}{\widetilde}
\newcommand{\wh}{\widehat}

\preprintnumber[3cm]{
YITP-99-1\\{\tt hep-th/9901001}\\January 1999}

\title{
String Junctions and Their Duals\\in Heterotic String Theory
}

\author{
Yosuke {\sc Imamura}\footnote{{\tt imamura@yukawa.kyoto-u.ac.jp}}
{}\footnote{
Supported in part by a Grant-in-Aid for Scientific
Research from the Ministry of Education, Science, Sports and Culture
(\#9110).}
}

\inst{
Yukawa Institute for Theoretical Physics,\\
Kyoto University, Kyoto 606-8502, Japan
}

\recdate{
\today
}

\abst{
We explicitly give the correspondence between spectra of heterotic string theory
compactified on $T^2$ and string junctions in type IIB theory compactified on $S^2$.
}

\begin{document}

\maketitle

\section{Introduction}

In type IIB string theory, there are many types of 7-branes labeled by two integers
$p$ and $q$, which specify the $SL(2,{\bf Z})$ monodromies around them.
If we go around a $[p,q]$ 7-brane in the counter-clockwise direction,
we obtain a $SL(2,{\bf Z})$ monodromy $M_{[p,q]}$ defined by
\begin{equation}
M_{[p,q]}
=\left(\begin{array}{cc}
 1-pq & p^2 \\
 -q^2 & 1+pq
 \end{array}\right),
\end{equation}
and to avoid multiple values of fields and string charges,
we should introduce branch cuts which give $SL(2,{\bf Z})$ transformations
compensating for the monodromies.
It was suggested by Zwiebach et al.\cite{fromopen,arbitrary,affine,uncovering,km,flavor} that
when 7-branes of a specific combination are on top of each other,
various groups containing exceptional groups, affine groups, and other exotic groups
appear as gauge groups on the 7-branes.
To construct $A_n$, $D_n$ and $E_n$ gauge theories,
we need three types of 7-branes
with labels $[1,0]$, $[1,-1]$ and $[1,1]$.
Following \cite{fromopen,arbitrary}, we call them A-, B- and C-branes, respectively.
The degrees of freedom contributing to the gauge symmetry enhancement arise from string junctions
connecting some of these 7-branes.
The $(p,q)$ charges of strings attached
on a 7-brane should be equal to the label of the 7-branes.

It is known that type IIB theory compactified on $S^2$,
which can also be expressed as a K3 compactification of F-theory,
is dual to heterotic string theory compactified on $T^2$.
On $S^2$ we have $24$ 7-branes
corresponding to singular fibers of elliptically fibered K3.
As is pointed out in \cite{km}, the junction lattice of the type IIB theory on $S^2$
has the signature $(+^{18},-^2)$, which is the same as the signature for the Narain lattice
of heterotic string theory.
Because there is unique $18+2$-dimensional Lorentzian self-dual even lattice,
it is trivial to show that the junction lattice
is identical to the Narain lattice on heterotic side.
The purpose of the present paper is, by means of dualities,
to establish the one to one correspondence
between invariant charges of junctions in type IIB theory
and the quantum numbers specifying
perturbative states of a heterotic string.

These two theories are dual in the manner we now describe.
On $S^2$, we have $24$ 7-branes.
Although there are many representations of these 7-branes,
we restrict ourselves to the case that they are represented as
four sets of $A^4BC$.
Once we obtain some results for this background,
we can extend them to other regions of moduli space
by a continuous deformation of the background.
In the weak coupling limit ($g_{\rm str}\rightarrow0$),
a pair consisting of a B-brane and a C-brane is reduced to an orientifold 7-plane.\cite{FandO}
Therefore, in this limit, we have four orientifold 7-planes,
and the compactification manifold $S^2$ becomes a $T^2/{\bf Z}_2$ orientifold.
By means of a T duality transformation along two directions on $T^2/{\bf Z}_2$,
the theory is transformed into type I theory on $T^2$,
and it is S dual to the $SO(32)$ heterotic string theory on $T^2$.

Through this series of duality transformations,
we can relate BPS states in heterotic string theory
to those in type IIB theory.
Let us assume the two compactified directions to be $x^8$ and $x^9$,
and call momenta and winding numbers along them $m_8'$, $m_9'$, $n_8$ and $n_9$.
(Generically, momenta $m_8'$ and $m_9'$ are not integers,
due to the presence of Wilson lines,
while winding numbers $n_8$ and $n_9$ are always integers.)
In Table \ref{dualcharge}, it is displayed how the states carrying these charges
are transformed by the dualities.
\begin{table}[htb]
\caption{Objects carrying charges $m_\mu'$ and $n_\mu$
in each theory related by dualities.}
\label{dualcharge}
\begin{center}
\begin{tabular}{ccccc}
\hline
\hline
         & $m_8'$   & $m_9'$   & $n_8$    & $n_9$    \\
\hline
hetero   & $x^8$ KK & $x^9$ KK & $x^8$ F1 & $x^9$ F1 \\
type I   & $x^8$ KK & $x^9$ KK & $x^8$ D1 & $x^9$ D1 \\
type IIA & $x^8$ F1 & $x^9$ KK & D0       & D2       \\
type IIB & $x^8$ F1 & $x^9$ F1 & $x^9$ D1 & $x^8$ D1 \\
\hline
\end{tabular}
\end{center}
\end{table}
As is known from this table,
charges $(P_\mu,Q_\mu)$ ($\mu=8,9$) of strings winding along the $x^8$ and $x^9$ directions
in type IIB theory are
represented by quantum numbers of heterotic string as follows:
\begin{equation}
P_8=m_8',\quad Q_8=-\frac{1}{2}n_9,\quad
P_9=m_9',\quad Q_9=\frac{1}{2}n_8.
\label{chargerel}
\end{equation}
The factor $1/2$ on the right-hand side
of the second and the fourth equation in (\ref{chargerel})
is a consequence of the fact that
the T dual of a winding type I D-string is half of a type IIA D-particle
on an orientifold 8-plane.

We should note that in generic 7-brane backgrounds,
there is a certain subtlety concerning the definition of winding numbers
of strings on the type IIB side.
We explain this in Section \ref{sec:correspondence}
before giving the final result.

\section{Spectrum of a Heterotic String}

If we assume the left mover (without tilde) is supersymmetric and
the right mover (with tilde) is bosonic,
the mass formula and the level matching condition for
$T^d$ compactified heterotic string theory are
given by
\begin{equation}
\frac{M^2}{8\pi T_{\rm str}}
=\frac{1}{2}(a_0^\mu)^2+(N_{\rm osc}+a)
=\frac{1}{2}(\wt a_0^\mu)^2+\frac{1}{2}(\wt{\bf a}_0)^2+(\wt N_{\rm osc}-1),
\label{massandlevel}
\end{equation}
where $\mu=10-d,\ldots,9$ labels the compactified directions,
and the constant $a$ is $-1/2$ for the NS sector and $0$ for the R sector.
The variables $a_0^\mu$ and $\wt a_0^\mu$
represent the zero mode of the compactified coordinates $X^\mu$,
and the vector $\wt{\bf a}_0$ is the zero-mode associated with
the 16-dimensional internal space of the right mover.
Due to the quantization of the momenta and wrapping numbers,
the $2d+16$-dimensional vector $(a_0^\mu,\wt a_0^\mu,\wt{\bf a}_0)$
takes values in the Lorentzian self-dual even lattice, the so-called `Narain lattice'.
For simplicity, let us assume the torus $T^d$ is rectangular.
Then, the zero modes are quantized as
\begin{equation}
\left(\begin{array}{c}
a^\mu_0 \\ \wt a^\mu_0 \\ \wt{\bf a}_0
\end{array}\right)
=
\left(\begin{array}{c}
\frac{l_s}{\sqrt2R_\mu}
     \left(m_\mu+{\bf w}_\mu\cdot{\bf k}
     -\frac{1}{2}{\bf w}_\mu\cdot{\bf w}_\nu n_\nu\right)
     -\frac{R_\mu}{\sqrt2l_s}n_\mu\\
\frac{l_s}{\sqrt2R_\mu}
     \left(m_\mu+{\bf w}_\mu\cdot{\bf k}
     -\frac{1}{2}{\bf w}_\mu\cdot{\bf w}_\nu n_\nu\right)
     +\frac{R_\mu}{\sqrt2l_s}n_\mu\\
{\bf k}-n_\mu{\bf w}_\mu
\end{array}\right),
\label{quantize}
\end{equation}
where $R_\mu$ are compactification radii,
$\bf k$ is a vector on a $16$-dimensional self-dual even lattice $\Gamma$, and
${\bf w}_\mu$ are Wilson lines which take values on the torus ${\bf R}^{16}/\Gamma$.
From this equation, we can read modified momenta $m_\mu'$,
which are momenta containing the shift due to the Wilson line,
as
\begin{equation}
m_\mu'=m_\mu+{\bf w}_\mu\cdot{\bf k}
     -\frac{1}{2}{\bf w}_\mu\cdot{\bf w}_\nu n_\nu,
\end{equation}
where $m_\mu$ and $n_\mu$ are integers.
Now, we are considering $SO(32)$ heterotic string theory,
and $\Gamma$ should be the $SO(32)$ weight lattice,
which is generated by the basis
\begin{equation}
\alpha_i={\bf e}_i-{\bf e}_{i+1} (i=1,\ldots,15),\quad
\alpha_{16}={\bf e}_{15}+{\bf e}_{16},\quad
\alpha_s=\frac{1}{2}\sum_{i=1}^{16}{\bf e}_i,
\label{so32basis}
\end{equation}
where the ${\bf e}_i$ form an orthonormal basis of ${\bf R}^{16}$.
For the following argument, it is convenient
to define the $2d+16$-dimensional vector
\begin{equation}
{\bf K}=({\bf k},m_{10-d},n_{10-d},\ldots,m_8,n_8,m_9,n_9).
\end{equation}
Furthermore, we introduce vectors ${\bf W}_\mu$
that satisfy
\begin{equation}
a_0^\mu={\bf W}_\mu\cdot{\bf K},
\end{equation}
where the inner product is defined by using the metric
\begin{equation}
g_{MN}=\left(\begin{array}{cccccc}
           {\bf1}_{16} \\
            & & 1 \\
            & 1 \\
            & & & \ddots \\
            & & & & & 1 \\
            & & & & 1
       \end{array}\right).
\end{equation}
For example, in the $S^1$ compactified case, the vector ${\bf W}_9$ is given by
\begin{equation}
{\bf W}_9=\frac{l_s}{\sqrt2R_9}\left({\bf w}_9,
         -\frac{1}{2}{\bf w}_9^2-\frac{R_9^2}{l_s^2},1\right).
\end{equation}
The vectors ${\bf W}_\mu$ also satisfy the equation
\begin{equation}
{\bf W}_\mu\cdot{\bf W}_\nu=-\delta_{\mu\nu}.
\end{equation}
Therefore, they form an orthonormal basis of
the $d$-dimensional sub-manifold with a negative definite metric.
The moduli parameters of the heterotic string theory are specified by giving
this sub-manifold and the string coupling constant,
and this implies that the moduli space of the heterotic string compactified on $T^d$ is
\begin{equation}
\frac{SO(16+d,d)}{SO(16+d)\times SO(d)\times SO(16+d,d;{\bf Z})}\times{\bf R}^+,
\label{hetmoduli}
\end{equation}
where ${\bf R}^+$ is associated with the string coupling constant $g_{\rm str}$.

The BPS condition for a heterotic string is
\begin{equation}
N_{\rm osc}+a=0.
\label{heterobps}
\end{equation}
For BPS states, the norm of $\bf K$ should be smaller than $2$:
\begin{equation}
{\bf K}^2
=(\wt{\bf a}_0)^2+(\wt a_0^\mu)^2-(a_0^\mu)^2
=-2(\wt N_{\rm osc}+\wt a)\leq2.\label{levelmatch}
\end{equation}
Equation (\ref{levelmatch}) corresponds
to the BPS condition for junctions $({\bf J}\cdot{\bf J})\leq2$
given in \cite{const,Geo}.
(In this paper, we adopt a signature for the intersection number
that is opposite to that in Refs.\cite{arbitrary,uncovering}.)
The mass formula can be rewritten as
\begin{equation}
\frac{M^2}{4\pi T_{\rm str}}=({\bf W}_\mu\cdot{\bf K})^2.\label{massiswk}
\end{equation}
In particular, massless states should satisfy
\begin{equation}
{\bf W}_\mu\cdot{\bf K}=0.\label{massless}
\end{equation}
Equation (\ref{massless}) defines a subspace of codimension $d$.
Because all of the ${\bf W}_\mu$ have negative norms
and the whole space has signature $(+^{d+16},-^d)$,
this subspace has a positive definite metric.
Because the lattice on the subspace corresponds to massless gauge fields,
the sub-lattice is nothing but a root lattice of the gauge group.
Let us introduce the basis ${\bf U}_i$ on the sub-lattice
corresponding to the fundamental weight.
Then, any vector $\bf K$ can be expanded as
\begin{equation}
{\bf K}=a_i{\bf U}_i+c_\mu{\bf W}_\mu.
\label{expandk}
\end{equation}
The coefficients $a_i$ give the Dynkin label of the state
associated with the vector ${\bf K}$.
On the other hand, the coefficients $c_\mu$ give central charges of the state,
because the mass square of each state is given as the sum of the square of the $c_\mu$:
\begin{equation}
\frac{M^2}{4\pi T_{\rm str}}=\sum_\mu c_\mu^2.
\label{massfromc}
\end{equation}

When we compactify heterotic string theory on a torus $T^d$,
new $2d$ $U(1)$ fields appear from the metric and the NS-NS 2-form field,
and at a generic point in moduli space,
the gauge group is a subgroup of
$SO(32)\times U(1)^{2d}$.
For example,
if we compactify the $x^8$ direction and introduce the Wilson line
\begin{equation}
{\bf w}_8=\left(\left(\frac{1}{2}\right)^n,0^{16-n}\right),\quad
(n<8)
\label{w8}
\end{equation}
then gauge symmetry is broken to $SO(2n)\times SO(32-2n)\times U(1)^2$.
If we adjust the compactification radius to the value
specified by
\begin{equation}
\frac{1}{2}{\bf w}_8^2+\frac{R_8^2}{l_s^2}=1,
\label{enhancer}
\end{equation}
then some winding modes become massless, and the gauge symmetry $SO(2n)\times U(1)$
is enhanced to $E_{n+1}$.
We can confirm this by observing the sub-lattice given by (\ref{massless}).
\section{Correspondence of states}
\label{sec:correspondence}

\begin{figure}[p]
\centerline{
\put(145,42){${\bf\Omega}^0$}%
\put(146,27){$\bf\Sigma$}%
\put(55,-5){$\bf\delta$}%
\put(-15,65){$A^nBCBC$}%
\put(45,70){$\scriptsize\left(\hspace{-.5em}\begin{array}{c}
                      1\\0\end{array}\hspace{-.5em}\right)$}%
\put(65,70){$\scriptsize\left(\hspace{-.5em}\begin{array}{c}
                      0\\\frac{1}{8-n}\end{array}\hspace{-.5em}\right)$}%
\put(100,55){$\scriptsize\left(\hspace{-.5em}\begin{array}{c}
                      0\\1\end{array}\hspace{-.5em}\right)$}%
\put(100,15){$\scriptsize\left(\hspace{-.5em}\begin{array}{c}
                      1\\0\end{array}\hspace{-.5em}\right)$}%
\epsfbox{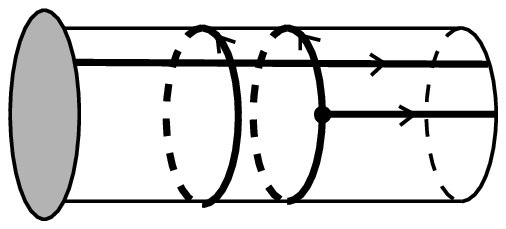}
}\VSPACE
\caption{Junctions ${\bf\Omega}^0$, $\bf\delta$ and $\bf\Sigma$.}
\label{ods.eps}
\end{figure}

In Ref.\cite{affine},
it is shown that the junction lattice associated with
the 7-brane background $A^nBCBC$ ($n<8$)
is identical to the root lattice of the affine Lie algebra $\wh E_{n+1}$.
On the heterotic side, this background corresponds to $T^2$ compactification
with the Wilson line (\ref{w8}) and the compactification radius $R_8$ satisfying (\ref{enhancer}).
In the 7-brane background,
in addition to junctions ${\bf\Omega}^i$ associated with the finite Lie algebra $E_{n+1}$,
we have three independent junctions, $\bf\delta$, ${\bf\Omega}^0$,
and $\bf\Sigma$ (Fig.\ref{ods.eps}).
An arbitrary junction on the background is expanded as follows:
\begin{equation}
{\bf J}=a_i{\bf\Omega}^i+k{\bf\Omega}^0+\wt n{\bf\delta}+\sigma{\bf\Sigma}.
\end{equation}
Coefficients $k$ and $\wt n$ correspond to the level and the grade, while
$\sigma$ is nothing but the asymptotic NS-NS charge.
The intersection numbers among these junctions are given in Table \ref{affinemetric}.
\begin{table}[hbt]
\caption{Intersection numbers among junctions
generating the affine Lie algebra $\wh E_{n+1}$}
\label{affinemetric}
\begin{displaymath}
\begin{array}{c|cccc}
               & {\bf\Omega}^i & {\bf\Omega}^0 & {\bf\delta} & {\bf\Sigma} \\
\hline
 {\bf\Omega}^i & A^{ij}       & & & \\
 {\bf\Omega}^0 &        & * & 1 & * \\
 {\bf\delta} &        & 1 & & \\
 {\bf\Sigma} &        & * & & -\frac{1}{8-n}
\end{array}
\end{displaymath}
\end{table}
Here $A^{ij}$ is the inverse matrix of the Cartan matrix of the finite Lie algebra $E_{n+1}$.
The components expressed as `$*$' in the table depend on the convention
used in defining the junctions ${\bf\Omega}^0$ and $\bf\Sigma$.
By adding $\bf\delta$ to them, we can fix these components to arbitrary values.
The correspondence of the vector ${\bf U}_i$ in (\ref{expandk})
and junctions ${\bf\Omega}^i$ is trivial.
Here, we focus on determining how we can reproduce the intersection $({\bf\Sigma}\cdot{\bf\Sigma})$,
which is independent of the convention.
Because the three junctions ${\bf\Omega}^0$, $\bf\delta$ and $\bf\Sigma$ carry central charges and
$\delta$ and $\Omega^0$ respect the same supersymmetry,
the contribution of these three charges to the mass
can be expressed as
\begin{equation}
\frac{M^2}{4\pi T_{\rm str}}=(xk+y\wt n)^2+z^2\sigma^2,
\end{equation}
where $x$, $y$ and $z$ are some constants.
If we assume the components `$*$' in Table \ref{affinemetric} are zero,
by comparing this equation with Eq.(\ref{massfromc}),
we obtain the following correspondence
between vectors on Narain lattice and junctions:
\begin{equation}
{\bf W}_8\sim z{\bf\Sigma},\quad
{\bf W}_9\sim y{\bf\Omega}^0+x{\bf\delta}.
\end{equation}
The problem is to determine how we should fix the coefficients $x$, $y$ and $z$.
For $z$, we can proceed as follows.
As is shown in Fig.\ref{ods.eps}, the junction $\bf\Sigma$ contains
a string loop with R-R charge $Q_9=1/(8-n)$.
By the relation (\ref{chargerel}),
we know that the configuration $\bf\Sigma$ on the type IIB side corresponds to
a heterotic string with winding number $n_8=2/(8-n)$.
On the other hand,
the $n_8$ component of the vector ${\bf W}_8$ is $l_s/(\sqrt2R_8)=2/\sqrt{8-n}$.
Therefore, we find
\begin{equation}
{\bf\Sigma}\sim\frac{1}{\sqrt{8-n}}{\bf W}_8.
\end{equation}
If we believe this correspondence,
we can reproduce the self-intersection number of $\bf\Sigma$ 
in Table \ref{affinemetric} as an inner product of the vector:
\begin{equation}
({\bf\Sigma}\cdot{\bf\Sigma})=\frac{1}{8-n}{\bf W}_8\cdot{\bf W}_8=-\frac{1}{8-n}.
\end{equation}

In this way, can we establish a complete correspondence between the quantum numbers
of heterotic strings and invariant charges of junctions?
Probably, this is possible.
However, it is difficult.
The origin of the difficulty is the existence of configurations like $\bf\Sigma$.
The position of the D-string loop can be freely moved,
and the length of the string attached to the loop depends on the position.
However, information regarding the position is not contained in the
quantum numbers on the heterotic string side.
Therefore it is not clear how we should define the winding number on the type IIB
side using heterotic quantum numbers.

\begin{figure}[hbt]
\centerline{
\put(80,150){$(P_8,Q_8)$}%
\put(80,120){$(P_8,Q_8)$}%
\put(15,80){$(P_9,Q_9)$}%
\put(70,80){$(P_9,Q_9)$}%
\epsfbox{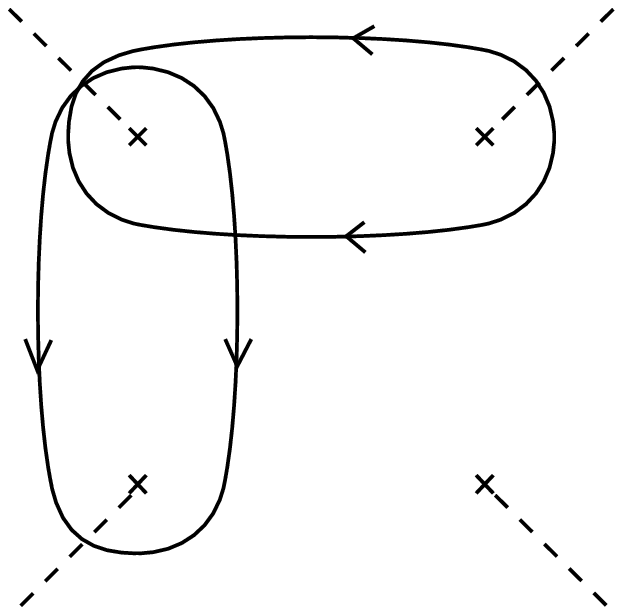}}\VSPACE
\caption{A dual configuration of a heterotic string
with Kaluza-Klein momentum $m_\mu'$ and winding number $n_\mu'$.
The string charges $(P_\mu,Q_\mu)$ are obtained from Eq.(\ref{chargerel}).}
\label{twoloops.eps}
\end{figure}

\begin{figure}[htb]
\centerline{
\put(65,190){$(2P_8,2Q_8)$}%
\put(190,85){$(2P_9,2Q_9)$}%
\put(90,130){$k_i-\frac{n_9}{2}$}%
\put(130,70){$k_i$}%
\put(25,70){$k_i-\frac{n_8}{2}$}%
\put(15,100){$k_i-\frac{n_8+n_9}{2}$}%
\epsfbox{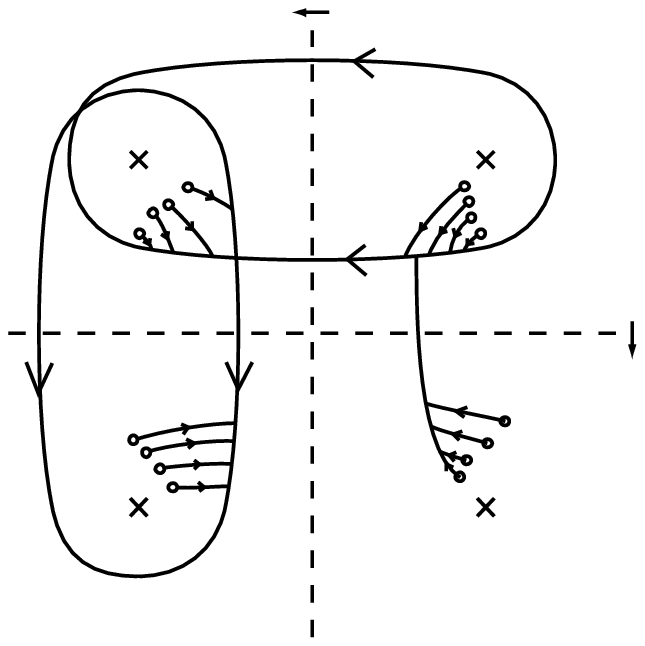}}\VSPACE
\caption{When $\wt{\bf a}_0$ does not vanish, we should define $(P_\mu,Q_\mu)$ as
one half of the string charges that go through $S^1$ cycles, expressed by the dashed lines.}
\label{onk3.eps}
\end{figure}

\begin{figure}[htb]
\centerline{
\put(100,-5){$A_1$}%
\put(115,10){$A_{16}$}%
\put(123,20){$B_1$}%
\put(127,33){$C_1$}%
\put(110,103){$B_2$}%
\put(100,115){$C_2$}%
\put(0,115){$B_3$}%
\put(-13,103){$C_3$}%
\put(-13,10){$B_4$}%
\put(2,-3){$C_4$}%
\epsfbox{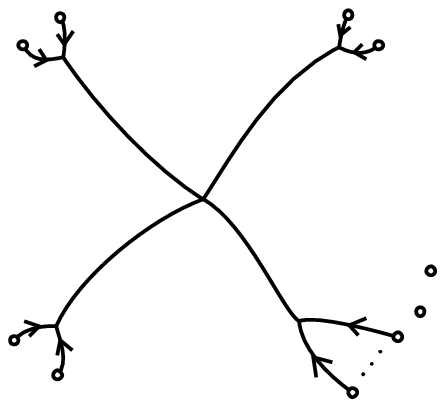}}\VSPACE
\caption{As a result of continuous deformation and
separation of O7-plane to B- and C-brane, we obtain a junction like this.}
\label{onk3d.eps}
\end{figure}

There is only one case in which such subtlety does not exist.
The reason we need the string attached on the string loop of the junction
$\bf\Sigma$ is that the monodromy around the loop changes the charge of the string loop.
Therefore, in the flat background, where the R-R charge of each O7-plane is canceled by
four D7-branes and the monodromy around the cycle is $1$,
configurations like $\bf\Sigma$ are not allowed,
and the definition of the winding number is very clear.
On the heterotic side, the background is specified by the following Wilson line:
\begin{equation}
{\bf w}_8=(0^8,(1/2)^8),\quad
{\bf w}_9=(0^4,(1/2)^4,0^4,(1/2)^4).
\end{equation}
On this background, there is no subtlety concerning the definition of charges,
and by means of Eq.(\ref{chargerel}),
we can represent the winding numbers of strings on the type IIB side
by quantum numbers on the heterotic side.
If all the components of the vector $\wt{\bf a}_0$ vanish,
we obtain a configuration with two string loops with charges $(P_\mu,Q_\mu)$
(Fig.\ref{twoloops.eps}).
If the components of the vector $\wt{\bf a}_0={\bf k}-n_8{\bf w}_8-n_9{\bf w}_9$
take nonzero values,
each component should be regarded as the number of strings attached to each D7-brane.
As a result, we obtain the configuration in Fig.\ref{onk3.eps}.
In this case, the charges of the string loops are not constant on them,
due to the fundamental strings lying between the loops and D7-branes.
Nevertheless, we can define the charge $(P_\mu,Q_\mu)$ of the junction uniquely
as one half of the sum of the charges
that go through cycles wrapped around the $x^8$ and $x^9$ directions,
which are represented by the dashed lines in Fig.\ref{onk3.eps}.
The charge of any segment in the configuration in Fig.\ref{onk3.eps} is
determined by the information we have already given.
In Fig.\ref{onk3.eps},
we assume that the branch cuts go outward.
By means of continuous deformation,
we can collect all $16$ D7-branes at the same position.
As we move D7-branes, the branch cuts attached on them are also moved.
If the branch cuts go across strings, the charges of the strings change,
and if a D7-brane goes across strings with D-string charge,
the number of strings attached to the D7-brane is
changed through the Hanany-Witten effect.
Taking account of these facts, we obtain the configuration in Fig.\ref{onk3d.eps}.
In this configuration,
we have separated each O7-plane into a B- and a C-brane,
and the $24$ 7-branes we have obtained are labeled
$A_1$, $\cdots$, $A_{16}$, $B_1$, $C_1$, $\cdots$, $B_4$, $C_4$
in counter-clockwise order.
As a result, invariant charges of the junctions are given as follows:
\begin{eqnarray}
Q_{{\bf a}_i}&=&k_i,\label{first}\\
Q_{{\bf b}_1}&=&0,\\
Q_{{\bf c}_1}&=&0,\\
Q_{{\bf b}_2}&=&-m_8+n_8-n_9-s,\\
Q_{{\bf c}_2}&=&-m_8+n_8-s,\\
Q_{{\bf b}_3}&=&m_8+m_9-n_8+s,\\
Q_{{\bf c}_3}&=&m_8+m_9-n_9+s,\\
Q_{{\bf b}_4}&=&-m_9+n_9-s,\\
Q_{{\bf c}_4}&=&-m_9-n_8+n_9-s,\label{last}
\end{eqnarray}
where $s$ is defined by
\begin{equation}
s=\frac{1}{2}\sum_{i=1}^{16}k_i.
\end{equation}
Because the vector $\bf k$ takes values on the lattice $\Gamma$,
$s$ is always an integer.

Junctions, which are dual to heterotic string states,
should be proper; i.e., their invariant charges should be integers.
At first sight, however, Eq.(\ref{first}) gives fractional charges
$Q_{{\bf a}_i}$ when $\bf k$ contains an odd number of spinor roots $\alpha_s$
in (\ref{so32basis}).
We can always make this junction proper
by adding null junctions.
We have two independent null junctions ${\bf N}_1$ and ${\bf N}_2$, which go around
all $24$ junctions.
Because the compactification manifold is $S^2$,
these junctions can shrink to a point.
They have the following invariant charges:
\begin{equation}
{\bf N}_1:(0^{16},1,1,-1,-1,1,1,-1,-1),\quad
{\bf N}_2:(1^{16},-15,-13,11,9,-7,-5,3,1).
\end{equation}
If all $Q_{{\bf a}_i}$ of a junction are half odd integer,
by adding the null junction $({\bf N}_1+{\bf N}_2)/2$,
we can make it a proper junction.
Furthermore, using null junctions,
we can fix any two of $Q_{{\bf b}_i}$ and $Q_{{\bf c}_i}$ to any values,
and in the relations (\ref{first})-(\ref{last})
we adopted the convention $Q_{{\bf b}_1}=Q_{{\bf c}_1}=0$.

Finally, let us confirm that the intersection number
defined in Ref.\cite{arbitrary} is consistent with
the inner product of the vector $\bf K$.
According to Ref.\cite{arbitrary},
the intersection numbers among unit junctions are
\begin{equation}
({\bf a}_i\cdot{\bf a}_j)
=({\bf b}_i\cdot{\bf b}_j)
=({\bf c}_i\cdot{\bf c}_j)=\delta_{ij},\quad
({\bf a}_i\cdot{\bf b}_j)=-\frac{1}{2},\quad
({\bf a}_i\cdot{\bf c}_j)=\frac{1}{2},\quad
\end{equation}
\begin{equation}
({\bf b}_i\cdot{\bf c}_j)
=\left\{\begin{array}{cc}
    -1 & (i\leq j)\\
     1 & (i>j).
 \end{array}\right.
\end{equation}
(In our convention, intersection numbers have signs opposite to those in Ref.\cite{arbitrary}.)
Using these equations, we obtain
\begin{equation}
({\bf J}\cdot{\bf J})={\bf k}^2+2m_8n_8+2m_9n_9.
\end{equation}
This is identical to the inner product ${\bf K}^2$ on the Narain lattice.
\section{Conclusion}
We have established the correspondence between the quantum numbers
of heterotic strings compactified on $T^2$ and invariant charges
of junctions on $S^2$ with $24$ 7-branes.
In order to find the necessary relations, we used a flat background on the type IIB side.
However, once we have obtained the relations (\ref{first})-(\ref{last}),
they are available for non-flat backgrounds,
because both the junction lattice and the Narain lattice are discrete,
and the relations between them
are invariant under continuous deformations
of moduli parameters.
In fact, the moduli space of elliptic K3 is
\begin{equation}
\frac{SO(18,2)}{SO(18)\times SO(2)\times SO(18,2;{\bf Z})}\times{\bf R}^+\times{\bf R}^+,
\label{K3moduli}
\end{equation}
where the two copies of ${\bf R}^+$ represent the K\"ahler structure
of the base manifold and that of the elliptic fiber, respectively.
In the case of the compactification of F-theory,
we should ignore the K\"ahler structure of the elliptic fiber.
Hence, the moduli space of type IIB theory compactified on $S^2$ is
identical to that of $T^2$-compactified
heterotic string theory, (\ref{hetmoduli}),
and it is expected that at any point on the moduli space
we can use the formulae (\ref{first})-(\ref{last}).
For example, applying these formulae to a vector ${\bf W}_8$
with the Wilson line (\ref{w8}),
we correctly obtain the junction $\bf\Sigma$, although the background is not flat.

\section*{Acknowledgements}

I would like to thank B.Zwiebach, K.Murakami and I.Kishimoto
for valuable discussions and B.Kol for a comment on the
moduli space of K3.


\end{document}